   \documentstyle[12pt]{article}
\newcommand\unlock{\catcode`\@=11}
\newcommand\lock{\catcode`\@=12}
\unlock
\def\maketitle{\par
 \begingroup
 \def\thefootnote{\fnsymbol{footnote}}
 \def\@makefnmark{\mbox{$^\@thefnmark$}}
 \@maketitle
 \@thanks
 \endgroup
 \setcounter{footnote}{0}
 \let\maketitle\relax
 \let\@maketitle\relax
 \gdef\@thanks{}\gdef\@author{}\gdef\@title{}\let\thanks\relax}
\def\@maketitle{\vspace*{0.9cm}
{\hsize\textwidth
 \linewidth\hsize \centering
 {\normalsize \bf \@title \par} \vskip 0.3cm  {\normalsize  \@author \par}}}

\def\thefootnote{\mbox{\noindent$\fnsymbol{footnote}$}}
    \long\def\@makefntext#1{\noindent$^{\@thefnmark}$#1}

\def\section{\@startsection{section}{1}{\z@}{1.5ex plus 0.5ex minus
   1.2ex}{1.3ex plus .1ex}{\normalsize\bf}}
\def\subsection{\@startsection{subsection}{2}{\z@}{1.5ex plus 0.5ex minus
    1.2ex}{1.3ex plus .1ex}{\normalsize\em}}

\def\@sect#1#2#3#4#5#6[#7]#8{\ifnum #2>\c@secnumdepth
     \def\@svsec{}\else
     \refstepcounter{#1}\edef\@svsec{\ifnum #2=1 \@sectname\fi
        \csname the#1\endcsname.\hskip 1em }\fi
     \@tempskipa #5\relax
      \ifdim \@tempskipa>\z@
        \begingroup #6\relax
          \@hangfrom{\hskip #3\relax\@svsec}{\interlinepenalty \@M #8\par}
        \endgroup
       \csname #1mark\endcsname{#7}\addcontentsline
         {toc}{#1}{\ifnum #2>\c@secnumdepth \else
                      \protect\numberline{\csname the#1\endcsname}\fi
                    #7}\else
        \def\@svsechd{#6\hskip #3\@svsec #8\csname #1mark\endcsname
                      {#7}\addcontentsline
                           {toc}{#1}{\ifnum #2>\c@secnumdepth \else
                             \protect\numberline{\csname the#1\endcsname}\fi
                       #7}}\fi
     \@xsect{#5}}

\def\@sectname{}

\def\thebibliography#1{\section*{{{\normalsize
\bf References }
\rule{0pt}{0pt}}\@mkboth
  {REFERENCES}{REFERENCES}}\list
  {{\arabic{enumi}.}}{\settowidth\labelwidth{{#1}}%
    \leftmargin\labelwidth  \frenchspacing
    \advance\leftmargin\labelsep
    \itemsep=-0.2cm
    \usecounter{enumi}}
    \def\newblock{\hskip .11em plus .33em minus -.07em}
    \sloppy
    \sfcode`\.=1000\relax}

\def\@cite#1#2{\unskip\nobreak\relax
    \def\@tempa{$\m@th^{\hbox{\the\scriptfont0 #1}}$}%
    \futurelet\@tempc\@citexx}
\def\@citexx{\ifx.\@tempc\let\@tempd=\@citepunct\else
    \ifx,\@tempc\let\@tempd=\@citepunct\else
    \let\@tempd=\@tempa\fi\fi\@tempd}
\def\@citepunct{\@tempc\edef\@sf{\spacefactor=\the\spacefactor\relax}\@tempa
    \@sf\@gobble}

\def\citenum#1{{\def\@cite##1##2{##1}\cite{#1}}}
\def\citea#1{\@cite{#1}{}}

\newcount\@tempcntc
\def\@citex[#1]#2{\if@filesw\immediate\write\@auxout{\string\citation{#2}}\fi
  \@tempcnta\z@\@tempcntb\m@ne\def\@citea{}\@cite{\@for\@citeb:=#2\do
    {\@ifundefined
       {b@\@citeb}{\@citeo\@tempcntb\m@ne\@citea\def\@citea{,}{\bf ?}\@warning
       {Citation `\@citeb' on page \thepage \space undefined}}%
    {\setbox\z@\hbox{\global\@tempcntc0\csname b@\@citeb\endcsname\relax}%
     \ifnum\@tempcntc=\z@ \@citeo\@tempcntb\m@ne
       \@citea\def\@citea{,}\hbox{\csname b@\@citeb\endcsname}%
     \else
      \advance\@tempcntb\@ne
      \ifnum\@tempcntb=\@tempcntc
      \else\advance\@tempcntb\m@ne\@citeo
      \@tempcnta\@tempcntc\@tempcntb\@tempcntc\fi\fi}}\@citeo}{#1}}
\def\@citeo{\ifnum\@tempcnta>\@tempcntb\else\@citea\def\@citea{,}%
  \ifnum\@tempcnta=\@tempcntb\the\@tempcnta\else
   {\advance\@tempcnta\@ne\ifnum\@tempcnta=\@tempcntb \else \def\@citea{--}\fi
    \advance\@tempcnta\m@ne\the\@tempcnta\@citea\the\@tempcntb}\fi\fi}

\def\abstract{\if@twocolumn
\section*{ABSTRACT}         
\else \small
\begin{center}
{ABSTRACT\vspace{-.5em}\vspace{0pt}}
\end{center}
\quotation
\fi}
\def\endabstract{\if@twocolumn\else\endquotation\fi}

\def\fnum@figure{Fig. \thefigure}

\long\def\@makecaption#1#2{
   \vskip 10pt
   \setbox\@tempboxa\hbox{\small #1. #2}
   \ifdim \wd\@tempboxa >\hsize    
      \small #1. #2\par            
   \else                           
      \hbox to\hsize{\hfil\box\@tempboxa\hfil}
   \fi}

\lock
   
\begin{document}
\hspace{4.7in} LBL-36296

\hspace{4.7in} October 1994
   \title{OUTREACH AND EDUCATION ON HIGH ENERGY PHYSICS}
   \author{R. MICHAEL BARNETT\thanks{\vrule height 18pt width 0pt This work
was supported by the Director, Office of Energy Research,
Office of High Energy and Nuclear Physics, Division of High Energy
Physics of the U. S. Department of Energy under Contract
DE-AC03-76SF00098, and by the U.S. National Science Foundation
under Agreement No. PHY-9320551.}\\
   {\em Physics Division, Lawrence Berkeley Laboratory, \\
   Berkeley,    CA 94720}}
   \maketitle
   \setlength{\baselineskip}{2.6ex}

\vglue 0.8cm
\begin{center}
{\tenrm ABSTRACT}
\end{center}
\vglue 0.3cm
{\rightskip=3pc
 \leftskip=3pc
 \tenrm\baselineskip=12pt
 \noindent
We review ongoing efforts and discuss possible future directions
in informing the public and educating students about Particle Physics.
 \vglue 0.6cm}
\baselineskip=14pt

\section{Introduction}

In the post-Cold War, post-SSC era, many of us in High Energy Physics
feel we need to redouble our efforts to communicate what we do and why
we do it.  While we believe that physicists should devote more time
to outreach and education, each of us feels that time is a scarcer resource
than money.  Nonetheless very substantial efforts are already underway,
and we just need to build upon these existing activities.  Ongoing efforts
range from those done by individuals, to those done by the labs and other
organized groups.  Inevitably there are questions about whether such efforts
are or should be done for selfish or altruistic motives.  I don't find
such discussions useful.  Many physicists have spent many years in
education and outreach; the public and our field have both benefitted.

The Division of Particles and Fields of the APS, as part of its current
study of future directions for our field, has created a working group
on ``Structural Issues.''  This working group formed a subgroup to
examine ongoing outreach and education efforts and seek proposals for
new directions.  It called the ``Public Outreach and Education Team'' (or
POET).
This group has had six activities so far:

\begin{itemize}

\item[1.]An evening meeting at Fermilab on February 14, 1994 at which general
issues were discussed.

 \vspace{-0.1in}

\item[2.]A survey of ongoing ideas and of proposals was conducted via e-mail.
The results of the survey and the Fermilab meeting were distributed via
e-mail.

 \vspace{-0.1in}

\item[3.]An electronic bulletin board was set for discussion of these issues
(but it has not been used much).

 \vspace{-0.1in}

\item[4.]At the Structural Issues Working Group meeting at LBL on July 9, 1994
a report was given and feedback obtained.

 \vspace{-0.1in}

\item[5.]At the DPF'94 meeting in Albuquerque on August 2, 1994, a plenary
session on Public Outreach and Education was held at which four reports
on these issues were given.

 \vspace{-0.1in}

\item[6.]At the DPF'94 meeting in Albuquerque on August 3, 1994, the
POET group met with conference attendees and had an intensive discussion
of proposals for future directions.

\end{itemize}

The DPF'94 meeting was the first general meeting at which a plenary session
was devoted to outreach and education in High Energy Physics.  Despite
the late hour of the session (4:30-6:00 PM) half of the attendees remained
for the entire session.  The moderator of the session was Geoffrey West (LANL).
The speakers included: Malcolm Browne (Pulitzer-Prize-winning writer for
the New York Times) on ``High Energy vs. Low Education: A National Challenge,''
Julia Thompson (Pittsburgh) on ``Outreach to Women
and Minorities,'' Ernest Malamud (SciTech and Fermilab) on ``Using Science
Centers to Expose the public to the Microworld.''  This report is a
summary of my talk in this session.

In general, we have heard many comments about the need to convey the
excitement of physics, and not just the big discoveries but the
controversies too.  There seems to be a consensus that we need to
find means to recognize and reward outreach and education activities
by physicists.  The holding of a plenary session at a major conference
was a good first step.

I am reporting here not on my own activities or ideas but on those
received through the survey and various meetings.
In the following sections, I will briefly review selected activities
and proposals concerning general lay audiences, students, the government,
the news media, etc.  The topics of targeted groups such as women and
minorities and of science museums were covered by other speakers.
I will discuss the use of the World Wide Web, the creation of a catalog,
and activities by collaborations.

Regrettably, the lack of space and
color make it impossible to reproduce the many color transparencies,
figures, and photographs that were sent to me.
Clearly many of us pursue physics education/outreach for physics as a whole,
but I will limit the scope of this report to HEP-related activities.

\section{Reaching Lay Audiences}

Many people have told us about their talks and classes for lay audiences.
In a later section I will discuss resources that are available to
people interested in presentations to any non-technical audience.
There are a variety of books available to the public about Particle
Physics topics.  There have been suggestions that someone should produce
a large-size ``coffee table'' photo book showing some of the detectors,
accelerators, events, and even people of High Energy Physics.

Two recent books brought to our attention were:
Cindy Schwarz's {\it A Tour of the
Subatomic Zoo}, which was written for the interested layperson/ undergraduate/
high school teacher or student.  It assumes no prior physics background
and can serve as an introduction to the basics. It was published by AIP.
Lawrence Krauss' {\it Fear of Physics}, tries to reach out to a
broad popular audience, in order to explain what physicists
are interested in and why.

The Florida State University Physics Department has been producing
mall exhibits for a number of years and report that they are quite popular.
CERN has set up its own science museum, MICROCOSM, and has now built a
separate building for it.  They say its purpose is ``to let the public see the
research work carried out by the physicists in their quest to
understand the laws of Nature.''  The number of visitors has remained
consistently very high.

\section{News Media}

In connection with the announcement of the initial top quark candidates,
Fermilab carried out an excellent program to inform the news media about
the physics and the experiments in a manner that allowed the media to report
the news accurately.  They put together a substantial package of information
well-suited for the target audience.  They had excellent results in getting
good coverage in many media outlets throughout the country.

We do not always have big news to report.  However, many people
have reported success at getting media coverage for aspects of their
experiments. Some sample comments:
``We had a press conference when we did the last touches on the experimental
hardware... \
It works, but takes a lot of work and courage... \
We were on the evening news. One key point was good contact to the public
relations office. One has to be extremely careful about the scope of
their press release... \
I think a press conference after the publication of key result or a
press conference when an experiment takes first data is the best
approach.''

One comment was ``It is often easy to interest journalists in
(well-defined) stories,
but it needs a significant effort to establish the `networking'
links to them.''  Some thought should be given to how we can do this.

A common problem felt by many physicists ``is the NEGATIVE peer pressure
to go public.'' The culture of our field has equated talking to the press
about one's research to ``publishing in the New York Times.''  Clearly
one should continue to publish in the standard journals, but in the
world we now live in, we are obligated to communicate our results,
our conclusions, and the benefits of our work to a broad audience.
Physicists should be {\it encouraged} to describe what we are doing and why.
We are excited by the theories and experiments of our field, and we
should not be ashamed to share that excitement.  These lessons have not
been lost on the astrophysicists; their stories appear weekly in the press
(even the less-glamorous stories).

A number of people at Fermilab have proposed a national meeting of science
writers and physicists to discuss the reporting of science.  Clearly
many of us feel that both the quantity and quality of reporting about
particle physics are not adequate.  It is a difficult subject about
which many writers may feel insecure.  Such a meeting might not only
help break down some these barriers, but would help foster contacts
between writers and physicists.

\section{Radio, TV, and Cable TV}

In general it is difficult to present science on television because
of the cost. However,
Bernice Durand (Wisconsin) teaches modern physics for nonscientists
very successfully on
Madison area cable TV where watchers know her as the ``physics lady.''

PBS has recently begun a new television series called
the {\it Magic School Bus}.  Several people has asked whether we
might be able to interest the producers in an episode on Particle
Physics.  It is a
fully animated children's educational series.
It features a teacher named Ms. Frizzle (played by Lily Tomlin),
who takes her students on a magically powered bus for scientific
field trips into the human body, around the solar system, or back
to the time of dinosaurs.
``Children's interest in science starts
to erode in the elementary grades," project organizers say.
{\it The Magic School Bus} project is designed to keep children's curiosity
alive.

People have noted that other sciences seem to be featured in 60-second
science profiles on the radio and have asked why not HEP.

\section{Government}

It is generally agreed that our field could do a better job of informing
Washington officials about what we do and why we do it, about the
benefits of our research, and about the excitement of particle physics.
Other areas of physics participate in APS' congressional visits programs
much more than we do. My own experience is that many Members of Congress
and their aides have never seen an HEP physicist and are happy for the
opportunity.  The recent Drell Panel report had a significant impact,
in part because of significant followup in Washington by members of the
panel and others.

It has been suggested that the DPF should sponsor occasional Congressional
Fellows similar to those from the APS and AAAS.  The cost is about
\$50,000 each in salary, moving expenses, etc. Unfortunately I doubt
that the DPF can afford this.  One should not underestimate the impact of
Congressional Fellows.  I have been told by aides in other offices
that these Fellows are regarded as ``gurus'' on science issues.
Unfortunately they have usually not been from our field, and in fact,
they have even campaigned {\it against} our interests.  It has been suggested
that we should simply push to end APS's program
which we pay for and which some believe may have done more harm than good with
respect to HEP.  These Fellows in no sense represent our field, nor is
it clear to some of our respondents that we get the best qualified
people to accept such positions.

A former Congressional aide has suggested that we would benefit more
by sponsoring quarterly receptions for Members and aides
(from the House and Senate) at which leading figures in HEP would discuss
HEP physics issues and developments.  He estimated that these evening
receptions would typically attract 15-20 people (assuming food was provided),
and felt that such numbers were well worthwhile.  This is already done by
other fields including chemistry and biological sciences.

\section{Science Community}

One of the lessons of the SSC debacle is that we could benefit from
better relations with the rest of the science community.  A British
correspondent reported that they have made great strides in improving
their relations with other communities and that it is greatly benefitted
them.

A proposal has been made to hold a meeting in Washington on the benefits
of basic research for America, cosponsoring it with biologists, chemists,
medical researchers, geologists, astronomers, etc.  Leading researchers
from each field would speak about the importance of basic research.
Reporters would be invited to attend.  Later participants in the meeting
could visit the Capitol to relate this message to whatever committees
or individuals are interested.  The purpose of the meeting would be
general and not to promote any particular projects.  It would serve the
dual purpose of reaching out to these other fields and explaining to
the public the value of basic research.

\section{Documenting the value of basic research in HEP}

A number of people have urged a new effort to document the impact
of basic research in areas ranging from education to technology
transfer to medical benefits to economic impact.  One suggestion is to
trace the history of particular technologies.  We have not received
any specific proposals on how to coordinate this.

\section{College students}

A recurrent theme from many respondents is that there are enormous numbers
of young people taking introductory physics courses in our own universities
and that we are wasting a tremendous opportunity by not turning them on
to physics and basic research as much as possible.  These people will
be the congressional aides, opinion leaders, etc. in a few years.

Others have proposed that we should spend more time giving talks at
neighboring colleges.

\section{Teachers and school children}

Many physicists are currently active in bringing particle physics
to high school students.  This can be done through presentations,
workshops, open houses, the creation of materials,  etc.  The national
laboratories all have such programs which I will discuss later.
One very active national group is the Contemporary Physics Education
Project (CPEP) which consists of teachers,educators, and physicists
(among the physicists are Cahn, Goldhaber, Quinn, Riordan, Schwarz, and
myself).
This group has created the wall chart on {\it Fundamental Particles
and Interactions} (in three sizes) and distributed more than 100,000 copies
of it.  It also has very popular color software for high school/college
students in both Mac and PC versions.  It mailed a packet of classroom
activities about particle physics to every high school physics teacher
in the US.  They are completing a book on the subject of particle physics,
detectors, accelerators, and astrophysics.  CPEP conducts many workshops
for teachers on how to use CPEP materials to teach particle physics.
CPEP has been featured in {\it Science}, {\it Physics Teacher}, and even
on the BBC World Service.

The American Chemical Society together with AIP periodically publishes
booklets for students with cartoons, etc.  The April 1993 issue was
on particle accelerators.  The book published by Cindy Schwarz with AIP
(described earlier) is intended for high school students.

A popular suggestion has been the idea of creating a catalog of
resources, materials, workshops, etc. on particle physics.
This would be made available (for free) not only to teachers but to physicists
to aid and stimulate them in joining education/outreach efforts.
The catalog would be available both in printed form and on the World-Wide
Web.  Some people propose mailing it to all high school teachers,
but others feel that would not be useful.

A number of people are currently making presentations and giving workshops
at teachers meetings such as the American Association of Physics Teachers
(AAPT) and the National Science Teachers Association (NSTA).  These
organizations have national, regional, and state meetings.  Those involved
in these presentations find them well received
and advocate that more people do it.

Another proposal is that we set up a national referral service (via
telephone and e-mail) that would direct high school and college
teachers with HEP questions to physicists who are willing to answer questions.
The idea would be to refer the teachers to physicists in or near their
own state.  They might call a number such as 1-800-PARTICLE (extra digits are
ignored).  This service may also provide a list of speakers.

Finally, physicists can and are working with local school districts
and state agencies.  In addition, there are university, college, high school
alliance programs (organized via the APS).

\section{Resources available to physicists}

Many of the national laboratories such as Fermilab, SLAC, Brookhaven, and
CEBAF
have substantial education departments that sponsor workshops and programs
for both students and teachers, and material development.  They are anxious
for the involvement of additional physicists.

Fermilab opened
the Leon M. Lederman Science Education Center in September 1992.  They have
their own building with many exhibits.  They have 45 precollege programs
serving over 40,000 teachers and 8,000 teachers per year.  In addition they
have many college programs.  They sponsor workshops for Latin American
countries and create Spanish versions of instructional materials.
Physicists are involved in Fermilab programs as research mentors,
seminar speakers, role models, question \& answer sessions with school
kids, consultants on science content, hands-on-science in the classroom,
museum volunteers, and SBIR proposals.  CEBAF programs emphasize ``Teach
science by doing science.''

Existing materials include transparencies, slides, comics, software, etc.
These will be included in the catalog discussed above in the section on
teachers and school children.
The public relations staff at laboratories and universities often have
resources available for physicists.

We should continue to report on outreach/education at DPF meetings
to inform physicists about resources and ongoing activities.  Many
have suggested that we should work through the DPF and other organizations.
We can also communicate about these activities via Internet bulletin boards
and newsgroups.

\section{Using the Information Superhighway}

More and more public schools are gaining access to the Internet.
One suggestion is that the labs should set up files from which events pictures,
detector designs, accelerator pictures, etc. can be obtained by anonymous ftp.
These should be appropriately annotated.

The World-Wide Web (WWW) presents tremendous opportunities as use is growing by
300 percent a year.  Major news media are searching the Web for stories,
among them the New York Times.  Even the sheriff of Tulsa, Oklahoma has listed
Tulsa's most wanted criminals on the Web

An example of the impact of WWW can be seen in the
interest generated by LBL's ``Whole Frog'' link-up.  Users can examine
many three-dimensional images of the frog with or without skin, from any
angle.  Different organs can also be seen separately.  In half a year
160,000 users from 56 countries have connected to it \hfil\break
(http://george.lbl.gov/ITG.hm.pg.docs/dissect/info.html).

CERN organized a major WWW Workshop on Teaching and Learning with the Web
in May 1994.  They had speakers and participants from throughout Europe
but few from the US.

NASA has placed on WWW tremendous numbers of images from the
Hubble Space telescope and elsewhere including pictures of supernova,
comets, galaxies, planets, etc.  These are annotated and sometimes
very useful for education.  There are also a variety of animations.
A prime focus of NASA pages is always on hot and current topics.
They have coordinated the efforts of their many different labs and facilities.

Fermilab has made great strides in making a major presence on the Web
with some excellent educational pages and a coherent, organized approach.
They cover the physics, the detectors, the accelerators, the benefits,
and more.  I suggest you look at it.

Clearly HEP (like NASA) should have a coordinated approach to the Web
with a single homepage for the public that points to the labs and other
relevant sites.  This effort may require a meeting of the interested
groups.  This page should contain short items summarizing the current
excitement and controversies in particle physics and point to lab
and university homepages for more information.

Physicists may also need to make some effort to aid schools and libraries
getting onto WWW.  Many are already on the Web (even some elementary
schools classes have their own pages), but most are not.

Other suggested approaches are to create multimedia CD-ROM programs about
particle physics or even Nintendo-type games.

The AIP has an e-mail news service on physics education.  It summarizes
information on resources, national initiatives, outreach programs,
grants, publications, etc.  To subscribe to AIP's PEN, send an e-mail
message to \ listserv@aip.org\ .  Leave the ``Subject'' line blank.  In
the body of the message, enter the following command:  $<$add pen$>$.

\section{Outreach by experimental collaborations}

One suggestion is that experimental collaborations should be responsible for
creating WWW and ordinary printed materials about their experiment.
These should  describe the physics motivations of their experiment and
explain how the experiment might accomplish these goals.  There are people
who believe that any experimental collaboration that cannot explain these
basic concepts to the public should not be funded.

Several people have suggested that experimental collaborations can do
much more.  A very interesting proposal is one under which traditionally
non-research colleges (and possibly high schools) could become ``affiliates''
of experimental collaborations.
Arrangements would be made whereby they would ``participate''
in research activities.
Their work might involve a small
scale hardware study (table top) or a simulation study.
They would need computer time or the loan of some small hardware system
for a few months.

An incentive for these schools would be very important: some degree of
recognition of being part of the experiment.
The institution names might be listed
on scientific papers under the banner ``educational
institutions.''

One possibility with CDF or D0 data for a college  senior lab
experiment would be to do some data analysis and
event reconstruction for particles such as $Z$, $W$, and top.

Once such educational material is developed, it could be distributed
to other colleges. Later it might be distributed to
high-level high schools as a test.

Astrophysicists have already developed such a program, and it has been
very successful.  It is called ``Hands on Universe.''  The organizers feel
it gives high school teachers and their students the opportunity to
become collaborators on real scientific research.  The program provides them
with access to professional grade telescopes, analytical tools and the training
to use them.  It is currently delivered to high schools across the United
States.  Students can request telescope time to obtain images of
the moon, planets, galaxies, or supernovae.

The program recently made national news (ABC Nightly News, Associated Press,
etc.) when two 17-year-old juniors at a Pennsylvania high school while
searching for a galaxy photographed a supernova (1994I).  While they did not,
of course, recognize this, their photograph was the earliest one taken
and therefore quite valuable.  Both the publicity for science and the
impact on young people were also valuable.

\section{Conclusions}

There is no doubt that there are some exciting things happening in high energy
physics outreach and education, carried out by educators and by physicists.
However, the reality is that extremely few physicists spend any time at
all on these efforts. They heartily endorse these programs, but find that
they lack either the time or the inclination to join in.

This plenary session was an attempt to change attitudes, and we thank the
conference organizers for their precedent-making initiative.  It is important
to show by our actions that we value public awareness.  We should make
communication a priority and reward it.  We need a mechanism to make this
happen, and motivation for people to do it.

As the conference's summary speaker (Howard Georgi) said, we need to think of
speaking with the news media as a means of informing the public about the
impact of public money spent on high energy physics, and we need to
stop calling it ``publishing in the New York Times.''

Conference participants who attended our POET meeting seemed especially
interested in the following proposals:

\begin{itemize}
\item[1)] Create a catalog of HEP resources (materials, workshops, etc.)
for teachers and for physicists.  It would be printed and on the World-Wide
Web.
 \vspace{-0.1in}
\item[2)] Together with basic researchers from other fields, organize
a meeting in Washington on the impact and importance of basic research.
 \vspace{-0.1in}
\item[3)] Organize a unified approach to presenting Particle Physics on the
World-Wide Web, presenting the highlights and controversies of our field.
 \vspace{-0.1in}
\item[4)] Begin a program of educational affiliates of experimental
collaborations who would perform specially designed analysis or experiments.
 \vspace{-0.1in}
\item[5)] Find means to better inform Washington staff and officials about
HEP (quarterly receptions  at the Capitol, congressional fellows, etc.).
 \vspace{-0.1in}
\item[6)] Organize a national science writers meeting with physicists.
 \vspace{-0.1in}
\item[7)] Encourage more HEP participation in science museum programs
and find means to present our subject in museum-type settings.
\end{itemize}

For these and other efforts to succeed, the DPF needs to give them some
priority and provide vital organizational support.  Moral support is welcome,
but if we wish for outreach and education activities to progress,
meaningful action by the DPF would be more beneficial.

\end{document}